\documentclass[]{article}

\usepackage[margin=1in]{geometry}

\usepackage{graphicx}

\usepackage{algorithm}
\usepackage[noend]{algpseudocode}
\usepackage{bold-extra}

\usepackage{amsfonts}
\usepackage{amsmath}

\usepackage{listings}

\usepackage{subcaption}

\usepackage{longtable}

\newcommand{\vone}{Pv1}
\newcommand{\vtwo}{Pv2}
\newcommand{\regions}[4]{\ensuremath{[#1, #2; #3, #4]}}
\newcommand{\ijkl}{\regions{i}{j}{k}{l}}
\newcommand{\cp}[1]{\textsc{#1}}

\newcommand{\hc}{HC}
\newcommand{\shc}{SHC}
\newcommand{\stem}{SC}
\newcommand{\fsc}{FSC}

\title{RNAprofiling 2.0: Enhanced cluster analysis of structural 
ensembles\footnote{\copyright\ 2023. This manuscript version is made available under the CC-BY-NC-ND 4.0 license.}
}

\author{Forrest Hurley, University of North Carolina at Chapel Hill \\
Christine Heitsch, Georgia Institute of Technology}

\begin{document}

\maketitle

\abstract{
Understanding the base pairing of an RNA sequence provides insight into 
its molecular structure.
By mining suboptimal sampling data, RNAprofiling~1.0 identifies 
the dominant helices in low-energy secondary structures as features, 
organizes them into profiles which partition the Boltzmann sample, 
and highlights key similarities/differences among the most informative,
i.e.\ selected, profiles in a graphical format.
Version~2.0 enhances every step of this approach.
First, the featured substructures are expanded from helices to stems. 
Second, profile selection includes low-frequency
pairings similar to featured ones.
In conjunction, these updates extend the utility of the method to 
sequences up to length 600, as evaluated over a sizable dataset.
Third, relationships are visualized in a decision tree which 
highlights the most important structural differences. 
Finally, this cluster analysis is made accessible to experimental 
researchers in a portable format as an interactive webpage, permitting
a much greater understanding of trade-offs among different possible
base pairing combinations. 
}

\section{Introduction}

The method of sampling RNA secondary structures from the Boltzmann
distribution~\cite{mccaskill-90,ding-lawrence-03} 
under the nearest neighbor thermodynamic model (NNTM)~\cite{turner-mathews-10}
provides critical base pairing alternatives to the minimum free 
energy (MFE) configuration.
Such information can be essential to understanding how RNA
sequences fold --- and the functionality of these important molecules.
Yet, the power of ensemble analysis can only be realized 
by identifying the underlying patterns in a sufficiently large set 
of suboptimal structures.

RNAprofiling, or just profiling for short, refers to the overall
cluster analysis method that organizes and analyzes a collection 
of secondary structures according to a set of features.
It was developed~\cite{profiling}
to identify the dominant combinations of base pairing signals
in the Boltzmann ensemble.
RNAprofiling~1.0 (denoted here \vone) consistently
achieves high sample compression together with
low information loss on ``small'' sequences,
on the order of 100 nucleotides (nt).
We present here an updated version, RNAprofiling~2.0 (denoted \vtwo),
which can mine a stable, informative structural signal from 
Boltzmann samples of much longer sequences.

In contrast to other cluster analysis methods like 
Sfold~\cite{ding-chan-lawrence-04} and RNAshapes~\cite{steffen-etal-06},
\vtwo\ does not generate the sample to be analyzed.
Rather, it is available to leverage the ensemble analysis power 
of state-of-the-art software packages like 
RNAstructure~\cite{Reuter2010} and ViennaRNA~\cite{Lorenz2011}.
Hence, we demonstrate here that \vtwo\ will reliably report 
the high probability base pairing combinations for sequences up to 
600 nt.

We note that the signal from the Boltzmann ensemble at 
the substructural unit level, i.e.\ the features being considered,
remains strong well-beyond 1000 nt.  
However, the probability of different \emph{combinations} of these
units, i.e.\ their profiles, decays with sequence length.
Like prediction accuracy, this is a reflection of the NNTM itself,
and the sampling method employed.
Given a particular Boltzmann sample as input, \vtwo\ outputs 
high quality information in a useful quantity for further hypothesis
generation.

As described,
the content of that information is determined directly from the input sample.
When introduced~\cite{profiling}, it was established that RNAprofiling
provides complementary information to both Sfold and RNAshapes.
Moreover,
a thorough analysis~\cite{wire-review} compared the three, 
where \vone\ analyzed Boltzmann samples generated by GTfold~\cite{gtfold}.
It was found that all three improved over the MFE,
but there was no clear advantage among cluster analysis methods in
terms of base pair prediction accuracy.

In terms of efficiency,
for a sequence of length $\sim$600 nt,
\vtwo\ analyzes a Boltzmann sample of 10,000 structures in about
20 seconds, with the sample generation taking about 5 sec.
Shorter sequences and/or smaller samples take correspondingly less time
to analyze, e.g.\ about 2 sec to analyze a sequence $\sim$200 nt 
and sample of 1000 structures.
In contrast~\cite{wire-review}, Sfold takes about 25 seconds 
(sampling + analysis) at this 200 length/1K size scale, as does RNAshapes. 

Regardless of which cluster analysis method is used, there are 
two key points for experimentalists~\cite{wire-review}.
First, as well-known to the ribonomics community, 
prediction quality improves if more than one conformation is considered.
Second, the quality is substantially enhanced if the conformations 
are initially considered at lower granularity/higher abstraction.
This supports a multilayered approach to RNA secondary structure 
determination where an early computational step identifies 
critical structural differences
``to be vetted by further computational analysis, experimental testing,
and/or biological insight.''

The new version of RNAprofiling presented here
significantly enhances the method's ability to do just that.
The new code is freely available at
\texttt{github.com/gtDMMB/RNAprofilingV2}
under a GPLv2 license 
and can be run online through the 
\texttt{rnaprofiling.gatech.edu} website.

\section{Method}

\vtwo\ follows the same three general steps as \vone.
First, identify key substructural units as features.
Second, cluster secondary structures into profiles
based on their features, and select the most informative.
Third, visually highlight relationships among the selected profiles.
As described, \vtwo\ provides significant enhancements at each step.
Additionally, the profiling output is now made available as an 
interactive webpage which further facilitates compare/contrast across clusters. 

\subsection{Feature identification}

\vone\ introduced ``helix class'' as its structural unit.
A helix $(i,j,k)$ in an RNA secondary structure is 
a set of base pairs $\{(i,j),(i+1, j-1),\ldots,(i+k-1,j-k+1)\}$ where 
at least one of $\{i-1, j+1\}$ and of $\{i+k,j-k\}$ are single-stranded.
Under the NNTM, a hairpin loop must contain at least 3 nucleotides
which implies that $j - i - 2k \geq 2$.
A helix $(i,j,k)$ is maximal if the minimum hairpin length is respected
and neither $(i-1,j+1)$ nor $(i+k, j-k)$ are a canonical base pair.
A helix class (HC) consists of all helices which are a subset of 
the same maximal helix, and is denoted by that maximal $(i,j,k)$ triplet.

Profiling always begins by determining all HC present 
along with their frequency (i.e.\ estimated probability) 
in the sampled secondary structures.
When the observed HC are ordered by decreasing frequency, 
this yields a distribution with a  
long tail of low-probability base pairings.
The threshold at which to cut the tail is determined by maximizing the
average Shannon information entropy~\cite{profiling}.
This yields a relatively small set of selected helix classes (SHC).

\vone\ used SHC as features, and demonstrated this achieves
both high sample compression as well as low information loss.
\vtwo\ generalizes this approach by using SHC to generate ``stem classes''
as defined below.

Like a \hc, a stem class (\stem) represents sets of possible
base pairings.
It is built from \shc\ and denoted as \ijkl, 
where the intervals $i \leq x \leq i+k-1$ and $j-l+1 \leq y \leq j$ 
are the minimal regions such that all base pairs 
$(x,y)$ represented have their 5' end in the first 
and 3' one in the second. 
A ``trivial'' \stem\ consists of a single \shc\ with \regions{i}{j}{k}{k}.
The \stem\ frequency is the number of sampled structures 
where any of the base pairings from the constituent \shc\ are present.

For our purposes,
two (or more) helices are said to form a stem if they are 
(successively) separated by at most two single-stranded bases on either side.
In other words, if a secondary structures contains helices $(i,j,k)$ 
and $(i',j',k')$ with $i+k \leq  i' \leq i+k+2$ and $j-k -2 \leq j' \leq j-k$,
then they form a stem.
This extends to multiple distinct helices in succession, as long as the 
separation criteria is met.
The idea of a stem is motivated by the NNTM~\cite{turner-mathews-10}, 
which treats small internal loops 
(i.e.\ with sizes $1 \times 1$, $1 \times 2$/$2 \times 1$, and $2 \times 2$)
as special cases, in part to address noncanonical base pairings.

Generalizing to the sets of base pairs which form 
profiling's structural units,
two HC are considered stemmable if there exists a 
helix from each class which together could form a stem.
If for some reason 
a smaller, or larger gap size than 2, is desired, this can be changed by
the user.
We note that stemmability is a reflexive and symmetric 
mathematical relation on HC.
A stem class (\stem) is then defined to be 
the transitive closure of stemmable pairs of SHC,
yielding well-defined equivalence classes.

Observe that the \hc\ definition naturally satisfies transitivity
whereas it is imposed on stemmable pairs.
To confirm that \stem\ remain local substructural units, we 
consider two properties: length and width.
Both are defined precisely in Supplemental Material.
Length is an upper bound on the number of pairings possible from 
the \stem\ in one structure.
Width is a measure of the `spread' where two helices which form 
a stem would have a width of 1, 2, or 3 depending on the 
asymmetry of the small internal loop/bulge.
As will be shown, nearly all \stem\ have this width as well.

\subsection{Profile selection} 

The next step is to cluster the sampled structures into ``profiles'' 
determined by a common set of features.
To focus on the most informative combinations, 
a maximum average entropy threshold is again used to filter the low
probability tail.
This yields a relatively small set of selected profiles for further
consideration.

In addition to using \stem\ rather than \shc\ as the default structural
unit,
\vtwo\ updates the estimated probabilities prior to profile selection 
by considering low frequency pairings similar to the featured ones.
These augmented counts
are distinguished as ``fuzzy'' stem classes (\fsc).
Intuitively, if `enough' non-\shc\ pairings in a secondary structure
span the 5' and 3' sequence regions \ijkl, when expanded slightly,
then the corresponding \stem\
is added to the \fsc\ profile for that structure 
and the \fsc\ frequency is increased accordingly.
(Details in Supplemental Material.)

In \vtwo, the profiles are selected based on \fsc\ by default.
In this way, low frequency pairings which occupy the same sequence 
regions, approximately, as the featured ones are included
in the summary profile graph information.
As with stem versus helix classes, and the stem gap size,
this can be changed by the user.

\subsection{Relationship visualization}\label{subsec:tree}

The third and final step is to highlight similarities and 
differences across the selected profiles.
This compare/contrast visualization illuminates crucial differences 
between the most frequently occurring combinations of base pairings.
In \vone, the summary profile graph was based on the Hasse diagram for the 
selected profiles ordered under set inclusion.
\vtwo\ retains this option but also provides (by default) a complementary 
perspective via an augmented decision tree. 
An example is seen in Figure~\ref{fig:output}, 
and discussed further in Section~\ref{subsec:webpage}.
Details of the unsupervised tree building procedure 
are in Supplemental Material.
The advantage of decision trees over Hasse diagrams
is that the number of edges 
remains small, even at longer sequence lengths,
making it more comprehensible for the user.
The trade-off is that support for non-selected feature combinations 
(the ``intersection profiles'' from \vone)
can be spread among different branches.

\subsection{Interactive output}\label{subsec:webpage}

\begin{figure}[ht]
\centering
\includegraphics[width=.95\textwidth]{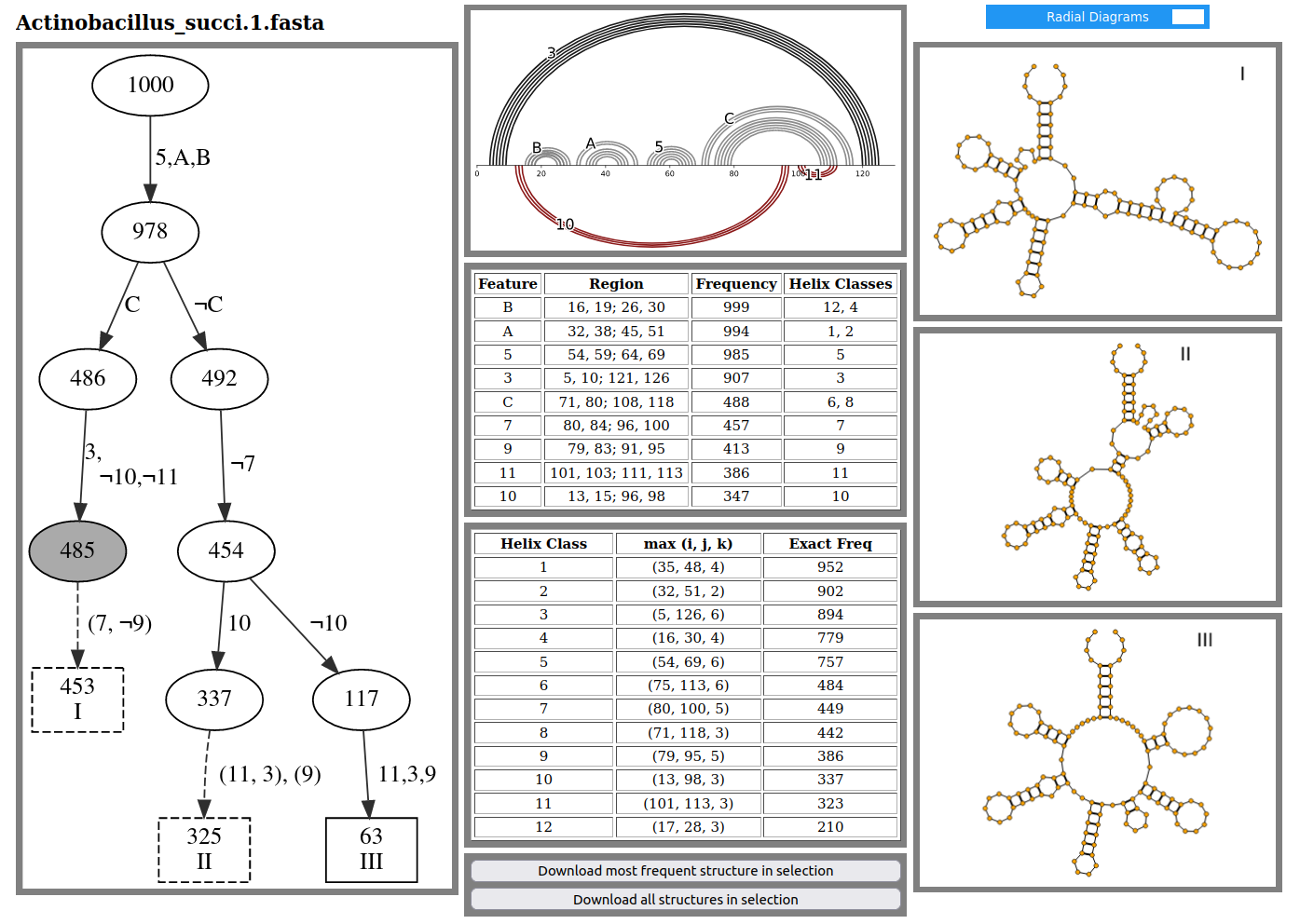}
\caption{\label{fig:output}
\small
Example of \vtwo\ output webpage.
Left column displays an interactive summary profile graph --- by default
a decision tree.
Nodes are clickable and labeled with corresponding number of sampled
structures.
Center column has a dynamic panel (top) displaying features 
in arc diagram format for chosen (grey) node from profile graph.
Decisions on incoming edge are emphasized; 
positive ones in bold above the sequence line,
and negative ones, denoted with $\neg$ in tree, 
in red below.
Features are labeled according to the \fsc\ table 
(center, middle) which lists \stem\ regions \ijkl\  
with fuzzy frequencies and \shc\ contained.
Nontrivial \fsc\ are denoted by letters, and trivial ones by their \shc\
index.
All \shc\ are listed (center, bottom) with maximal $(i,j,k)$ triplet 
and integer indexed in decreasing exact frequency as given.
Selected profiles, or groups thereof, are denoted by rectangular leaves
in decision tree and labeled by roman numerals. 
More than one selected profile is represented if the incoming edge is
a contingency (dashed).
In this case, the contingency table is given below the feature display
when the leaf (dashed rectangle) is chosen.
Right column shows most frequent secondary structure  
for each leaf in radial (or arc) diagram format.
Users can download all structures corresponding to the chosen node,
or just the most frequent, for further analysis.
See Section~\ref{subsec:webpage}, and Supplemental Material, 
for further information.
}
\end{figure}

The most visible enhancement in \vtwo\ is the profiling output; 
the summary profile graph is now embedded 
in a portable interactive webpage with auxiliary information.
Figure~\ref{fig:output} shows an example 
for the FMN riboswitch (RF00050)
in Actinobacillus succinogenes~130Z (CP000746.1/533105--533234).
As illustrated, this provides a compact, informative summary 
of the Boltzmann sample input
which also highlights critical structural differences.

Figure~\ref{fig:output} caption gives an overview of the information provided.
We now consider the decision tree example in more detail.
All nodes (ovals and rectangles) correspond 
to a subset of the Boltzmann sample input, and are labeled by its size.
Rectangles can only be a leaf, and denote a selected profile
or collection thereof (dashed).
The root node is the full sample provided.
Edges are labeled with choices on feature inclusion and/or exclusion
(denoted $\neg$) that lead to at least one selected profile.

As detailed in Supplemental Material, decisions are grouped based 
on the split induced by each feature on the selected profiles among 
the current set of structures.
If there is no forced decision, the one maximizing the dissimilarity
in the two branches (according to the Hellinger distance~\cite{cam-yang-00})
is chosen.
Splits indicate critical structural differences, in order of priority 
for further computational analysis and/or experimental testing.

For instance, of the full 1000 structures in the sample analyzed
for Figure~\ref{fig:output},
978 contain base pairings from 5, A, and B.  
For these structures, all selected profiles contain all three features, 
so this is a ``forced'' decision.

Letter indexed features denote nontrivial \fsc.
As seen in the feature display, and confirmed in the 
\fsc\ and \shc\ tables,
feature A is composed of two \shc, 1 and 2, that are 
separated by a small internal loop of size 
1 $\times$ 1, which could be a noncanonical pairing.
In contrast, B contains 4 = (16,30,4) and 12 = (17,28,3)
which can form a stem with 4 base pairs and a bulge of size 1.
Hence, \vtwo\ treats them as a single structural unit.

Fuzziness increases the estimated probability of these \stem\ 
by less than 5\% (from exact frequencies of 952 for A and 954 for B).
However, it raises 5 by more than 30\%, from 757 to 985.
This is useful information that the sequence region [54,59; 64,69] 
most likely forms some kind of hairpin structure.
None of the other combinations of inclusion/exclusion on 
5, A, B lead to a selected profile, 
so there is only one edge in the tree for this common forced decision.

In contrast, the 979 structures are essentially evenly split between 
including/excluding C.
According to the Hellinger distance,
this makes it the highest priority structural difference among the 
remaining features for this Boltzmann sample input.

If C is present, then this forces
3 to be included, 10 and 11 to be excluded.
Why exclusions are forced is useful information, and communicated 
by red pairings below the sequence line in the feature display.
In this case, the only remaining feature uncertainty (as summarized
by a contingency table) is between 7 and 9.
Given the overlap in their base pairing regions, 
resolving this ambiguity may not even be necessary. 

If C is absent, the next critical split identified is 10.
If it exists, then different combinations of the inclusion/exclusion 
of 9 and of (11,3) are present in the sample,
as would be summarized in the contingency table for II. 
The table would show that of the 325 structures on this decision path,
most (0.575) have both but some (0.169) have 9 with
($\neg$11, $\neg$3) 
while others (0.178) have the opposite (and 0.077 have neither).

Finally, excluding both C and 10 yields a single selected profile III 
pictured in bottom right.
Based on this analysis, the most important difference between the 
structures in the input sample
is the presence of C, and if $\neg$C then 10.
To facilitate further investigation,
after selecting a node, users can download files containing all associated
structures.
All images in the webpage can be downloaded directly or are available 
in the output folder.

\section{Results}

Previous results~\cite{profiling} demonstrated that 
\vone\ consistently 
achieves high sample compression together 
with low information loss on ``small'' sequences 
around 100 nt.
We confirm this on a much larger dataset, and also demonstrate that
the \vtwo\ enhancements extend the length
at which a useful structural signal can be extracted 
from a Boltzmann sample up to 600 nt.
Results highlight the value of this cluster analysis 
approach, 
as well as the challenges for very long sequences.

\subsection{Dataset}

Our dataset used the curated  
CONTRAfold~\cite{Do2006} one, which 
contains 151 sequences from distinct Rfam~\cite{Kalvari2021}
families (version 7.0), as a starting point.
Each existing family was augmented by up to 4 additional sequences 
from Rfam (version 14.9).
Eight of the original families are no longer in Rfam, while
3 had just one more sequence, 8 had two, and 7 had 3.
For the remaining 125 families, \cp{g}\cp{c} content was used 
as a proxy for phylogenetic diversity; 
four sequences were added iteratively by maximizing the minimum 
difference with the previous ones. 
This yielded a total of 683 sequences over 143 current families.

For each sequence, 25 different Boltzmann samples of size $10,000$
were generated using the RNAlib python bindings~\cite{Lorenz2011}.
Analysis reports the results of these 17,075 trials.
As addressed below, larger sample sizes 
(beyond the typical 1000~\cite{ding-lawrence-03})
improved result reproducibility
across different runs for longer sequences.

The maximum CONTRAfold sequence length was 568.
Excepting one family (RF00177), all additional ones
have length $< 600$\ nt.
To facilitate profile-level comparisons, the families were divided 
by average sequence length $n$
into 5 categories: 
extra-small (\cp{xs}), small (\cp{s}), medium (\cp{m}), 
large (\cp{l}), and extra-large (\cp{xl}).
They consist, respectively, of
$14$ families with $23 < n \leq 50$,
$85$ with $50 < n \leq 150$, $22$ with $150 < n \leq 220$, 
$21$ with $270 < n \leq 567$, and $1$ with $n = 1331.2$.

The \vone\ proof-of-principle dataset had 15 sequences spread
over 5 families with lengths ranging from 72 (tRNA) to 133 (5S rRNA) 
with a 99 nt average.
As shown below, \shc\ consistently yield profiles with high
sample compression and low information loss over the 85 comparable
\cp{s} families.
Moreover, by moving to \fsc\ as features, \vtwo\ achieves the same profile 
qualities on \cp{m} families,
and even extracts a useful signal for \cp{l} ones.
The \cp{xl} outlier was retained as an example 
of how much the structural signal in a Boltzmann sample,
i.e.\ the combination of base pairings, decays
with sequence length.

\subsection{Features}

\vone\ results found that Boltzmann sampling can be very noisy, 
with many distinct helices generated even at small lengths.
This is only more true for longer sequences with larger (by $10\times$)
sample sizes.
Nonetheless, relatively few features can 
reproducibly represent most of the base pairing information in 
a Boltzmann sample.

We found linear relationships between sequence length and  
number of different features.  
(See Supplemental Material.)
There are about $500$ distinct helices per $100$ nt on average,
and moving to HCs yields a 2-fold reduction. 
The maximum average entropy thresholding achieves about 25-fold reduction,
with $10$ SHC per 100 nt.
This is further reduced to a rate of $5.14$ \stem\ $(R^2 = 0.96)$. 
Since fuzzy counts only affect the estimated probability,
the number of distinct \stem/\fsc\ per 100 nt 
is two orders of magnitude less than helices.

This amount of compression still provides good coverage of
the base pairing information sampled.
Coverage was computed as the proportion of helices (with multiplicity)
which belong to a \shc. 
Nearly half (48.4\%) of trials had $> 0.90$ coverage, and most 
(87.3\%) covered more than 3/4 of helices sampled.
Coverage decreased slightly with sequence length (at a rate of 0.025 
per 100 nt according to the best line fit).
Only 2.0\% had coverage below $0.6$, and these outliers were 
generally shorter sequences ($< 150$ nt).

The \stem\ are built from \shc\ so provide the same coverage with
1/2 as many structural units.
We confirm that \stem\ remain local substructures by considering 
the length and width distributions.
As defined, a helix has width $w = 1$, and two helices that form a stem
have $1 \leq w \leq 3$  depending on the asymmetry of the 
small internal loop/bulge between them.
Over all trials, more than half ($52.2\%$) contained only
one \shc, and these trivial \stem\
had average (std) length $l = 5.45\ (2.39)$ nt.  
Allowing stems increases the feature length 
without significantly increasing the width.
Of the nontrivial \stem, nearly all ($89.1\%$) had $w \leq 3$ with 
$l = 11.9 \ (5.21)$ 
while only $1.2\%$ had $w \geq 6$ with $l = 28.7\ (18.6)$.

The reproducibility of \shc\ is very high, across all lengths.
For each sequence, this is the average over all \shc\ sampled
of the percentage of runs (out of 25) that it is present.
See Supplemental Material for boxplots. 
Reproducibility for \stem, as well as profiles, is computed likewise. 
However, the criteria for ``is present'' is stringent;
the compound structure must include exactly the same set of \shc.
This means that any variation in \shc\ propagates and is amplified.
Nonetheless, median \stem\ reproducibility remained high.
Interestingly, the noisier sequences were the shorter ones.

Boltzmann samples
of size 1K were originally analyzed, and only 10 (of 683) sequences 
had an average \shc\ reproducibility $h \leq 0.9$.  
Hence, more than 3/4 ($76.9\%$) had an 
average \stem\ reproducibility 
$s > 0.9$.
Moving to size 10K, as will be discussed in the next section,
meant only one (\cp{xs}) sequence had $h \leq 0.9$ which improved 
the percentage with $s > 0.9$ to $93.0\%$.
This included all but 4 (of 106) \cp{m} sequences and 1 (of 105) \cp{l} ones,
which had $0.8 < s \leq 0.9$.

This confirms that \vtwo\ is reliably extracting a very informative
base pairing signal from the sampled structures.  
Moreover, since the number of possible profiles grows exponentially with 
features, the greater compression obtained by \vtwo\
makes this approach accessible to longer sequences.

\subsection{Profiles}

Profiles were also assessed for sample compression, information loss,
and reproducibility.  
Five different feature sets were considered: 
helix, \hc, \shc, \stem, and \fsc.
For each type, the corresponding profiles were generated,
and the maximum average entropy thresholding applied to select
the most informative ones.
Results are broken down into groups by average sequence length per family,
and report analysis of 10K-sized Boltzmann samples except where noted.

\begin{figure}
\centering
\includegraphics[width=.95\textwidth]{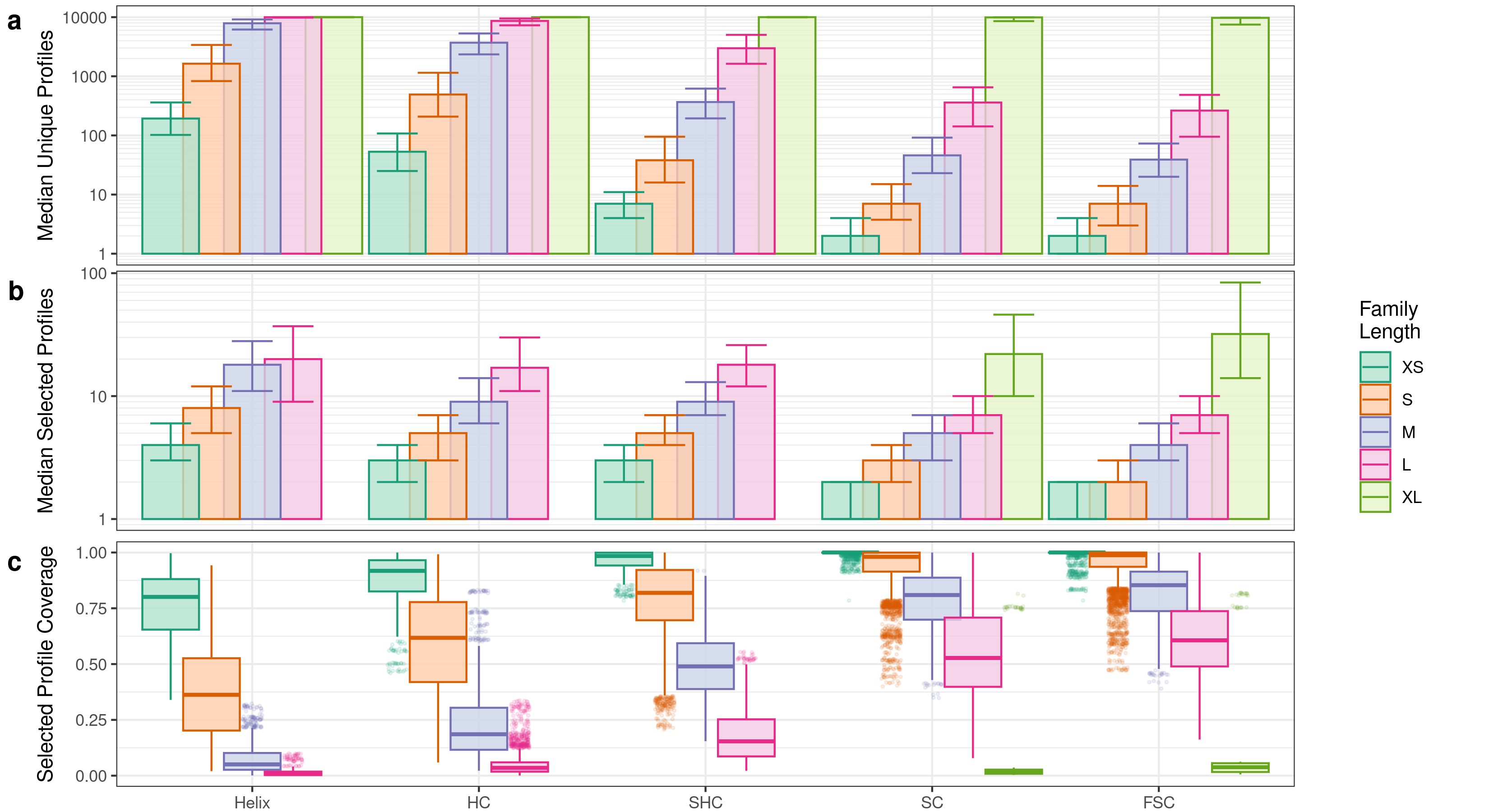}
\caption{\label{fig:profiles}\small
Profile analysis with feature sets ordered by 
increasing abstraction/decreasing granularity from helices through 
helix classes to stem classes.
Median and interquartile range (IQR) are reported over distinct trials,
excluding those where every sampled structure is a unique profile.
Counts (top and middle) are log scale.
Coverage (bottom) is proportion of sampled structures represented by 
selected profiles.
}
\end{figure}

Results given in Figure~\ref{fig:profiles} demonstrates that \vtwo\
achieves high sample compression with low information loss for 
sequences up to 600 nt.
Using \shc\ as features works well in the length range (up to 150 nt)
previously reported.
The number of selected profiles is low and the coverage is high:
3 and 0.98 for \cp{xs}, 5 and 0.82 for \cp{s}. 
However, the coverage for \cp{m} has already dropped to 0.49.
The key to 
overcoming this length barrier is consolidating \shc\ into \stem. 

As seen, Boltzmann sampling can be very noisy.
Even for the \cp{s} sequences,
a median of 1,634 different structures (i.e.\ helix profiles) were sampled.
Decreasing feature granularity filters sample noise.
The difference is particularly noticeable for \cp{l} sequences when
moving from \shc\ to \stem; the median drops more than 8-fold
(from 2,984 to 360).
Moreover, the median number of selected profiles is reduced down to 7,
not much higher than \cp{m} and \cp{s}.
Importantly, this level of sample compression is accompanied by a
corresponding rise in median coverage,
which increases from 0.15 to 0.61 for \fsc.
The corresponding rise for \cp{m} from 0.49 was to 0.85.

When analyzing reproducibility,
the standard 1K-sized samples~\cite{ding-chan-lawrence-04}
did not yield results comparable to
\vone, but 10K did.
Profile reproducibility is affected by the propagation, and 
amplification, of feature variability;
increased sample size reduces this effect across all sequence lengths.
For example, with 1K samples, although $80.2\%$ of \cp{m} sequence
have an average \shc\ selected profile reproducibility $s > 0.7$,
only $10.4\%$ were above $0.9$.  
With 10K samples, $98.1\%$ have $s > 0.7$ with $69.4\%$ above $0.9$.
Although the corresponding numbers for \cp{l} sequences also improve
significantly
(to $84.8\%$ from $20.0\%$ and $18.1\%$ from $0\%$ resp.), 
they are lower due to the confounding effect of profile growth.

Like the number of structural units sampled,
the average number of features in a profile also grows linearly with  
sequence length 
(at a rate of about $6$ \shc\ and $4$ \stem/\fsc\ per 100 nt). 
This effect is particularly apparent for the \cp{xl} sequences, 
whose \shc\ and \stem\ reproducibility is as high as any other,
but whose corresponding profiles are not reproducible.
(See Supplemental Material.)

Since \vtwo\ uses \fsc\ 
by default, these are the selected profiles whose reproducibility 
was evaluated.
It is very high for \cp{xs} and \cp{s} sequences ($92.1\%$ and 
$73.5\%$ above $0.9$, resp.)
although the outliers in \stem\ reproducibility are apparent.
Nearly all \cp{m} sequences ($88.7\%$) are above $0.7$, with a majority
($57.1\%$) above $0.9$.
Even the \cp{l} ones have $64.8\%$ and $35.2\%$ resp., with a median 
of $0.84$ and interquartile range of $(0.61,0.93)$.

\section{Conclusions}

RNAprofiling~2.0 (\vtwo) consistently achieves high sample compression together
with low information loss on sequences up to 600 nt,
a 4-fold length increase over \vone.
This is accomplished by expanding the featured substructures from helices
to stems, including low-frequency pairings similar to featured ones
in the profile selection,
and visualizing their relationships in a decision tree.
\vtwo\ takes as input a Boltzmann sample, as provided by software
packages like RNAstructure or ViennaRNA.
The \vtwo\ output is a portable interactive webpage which provides 
a compact, informative summary of the sample provided.
Critical structural differences are highlighted to be evaluated further
by some combination of computational analysis, experimental 
testing, and biological insight.

\section{Acknowledgments}

This work was supported by funds from the National Institutes of Health 
(R01GM126554 to CH) and the National Science Foundation (DMS1344199 to CH).
Additional support for FH provided by a grant from the
National Institute of Environmental Health Sciences
(2T32ES007018 to Rebecca Fry, UNC).

\end{document}


\title{Supplemental Material for \\
RNAprofiling 2.0: Enhanced cluster analysis of structural 
ensembles\footnote{\copyright\ 2023. This manuscript version is made available under the CC-BY-NC-ND 4.0 license.}}

\author{Forrest Hurley, University of North Carolina at Chapel Hill \\
Christine Heitsch, Georgia Institute of Technology}

\maketitle

\section{Additional results}

\begin{figure}
    \centering
	\includegraphics[width= \textwidth]{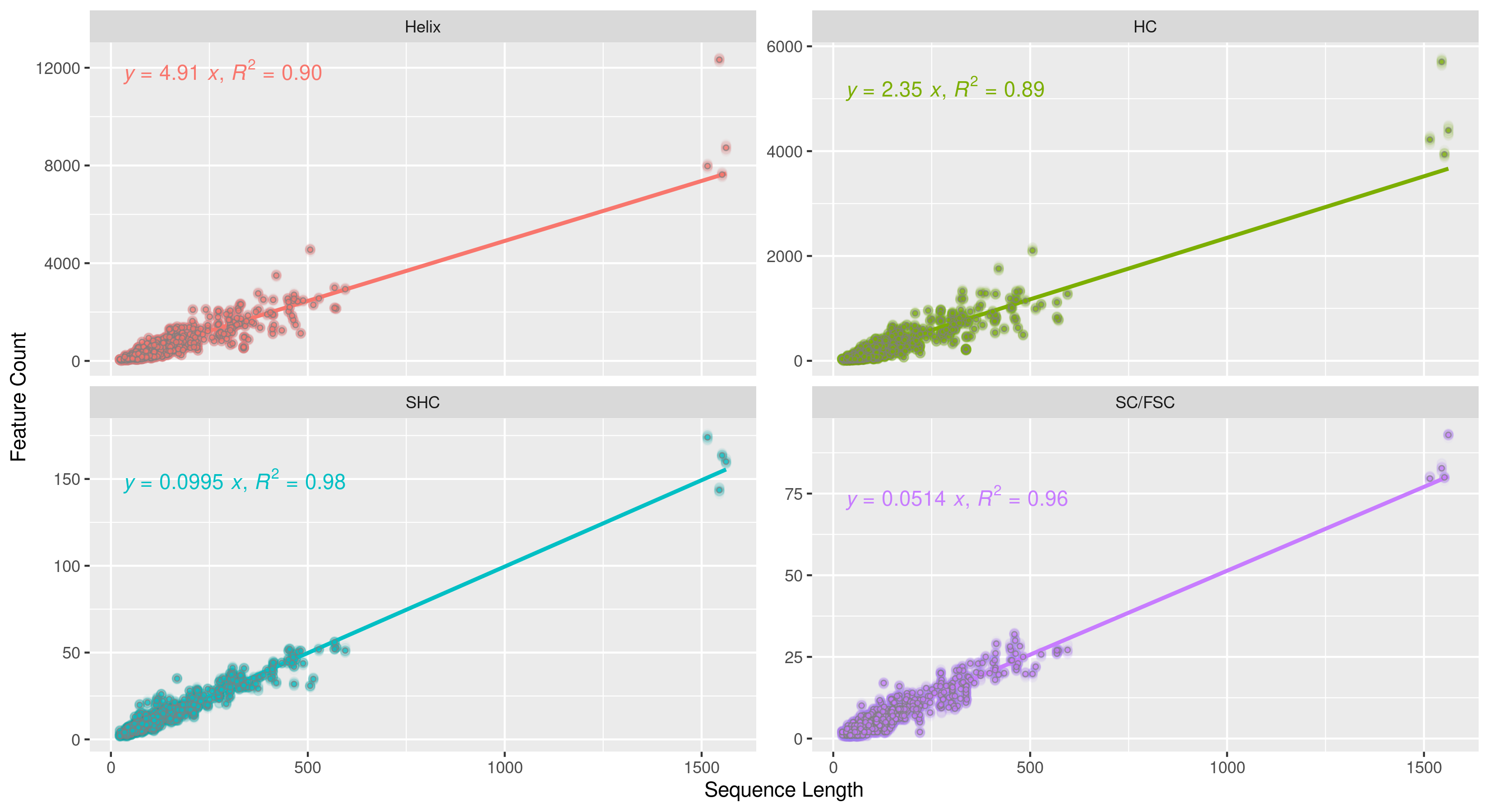}
    \caption{Number of distinct features by type for each sequence. 
Regression lines assume 0 intercept.
Note difference in y-axis scales between graphs.}
    \label{fig:supp_feature_count}
\end{figure}

\begin{figure}
    \centering
	\includegraphics[width= \textwidth]{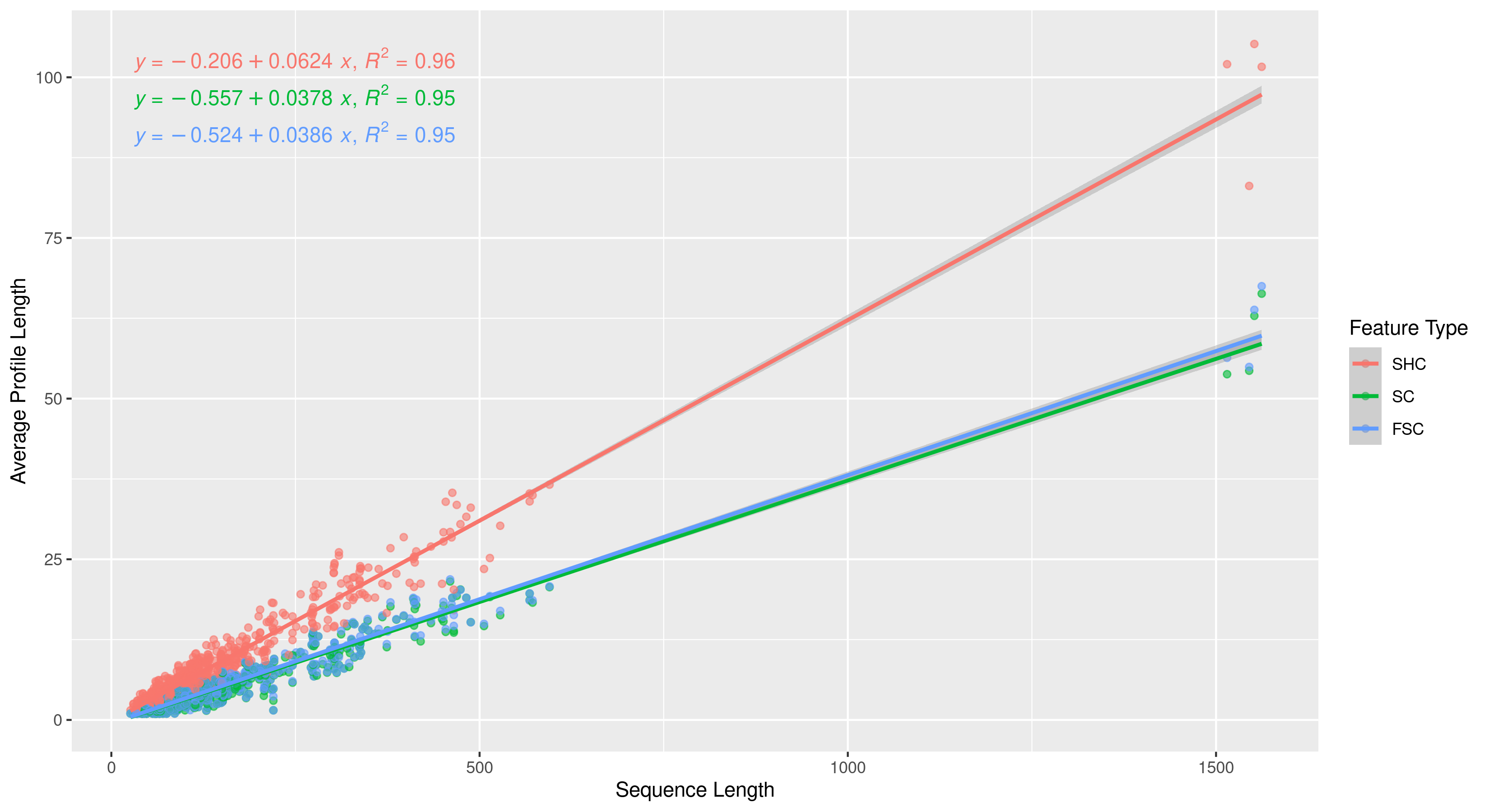}
    \caption{Average length of profiles by feature type for each sequence. 
Difference in regression line slopes 
	for \stem\ and \fsc\ is not significant (p=0.08).}
    \label{fig:supp_profile_length}
\end{figure}

\begin{figure}
    \centering
	\includegraphics[width= \textwidth]{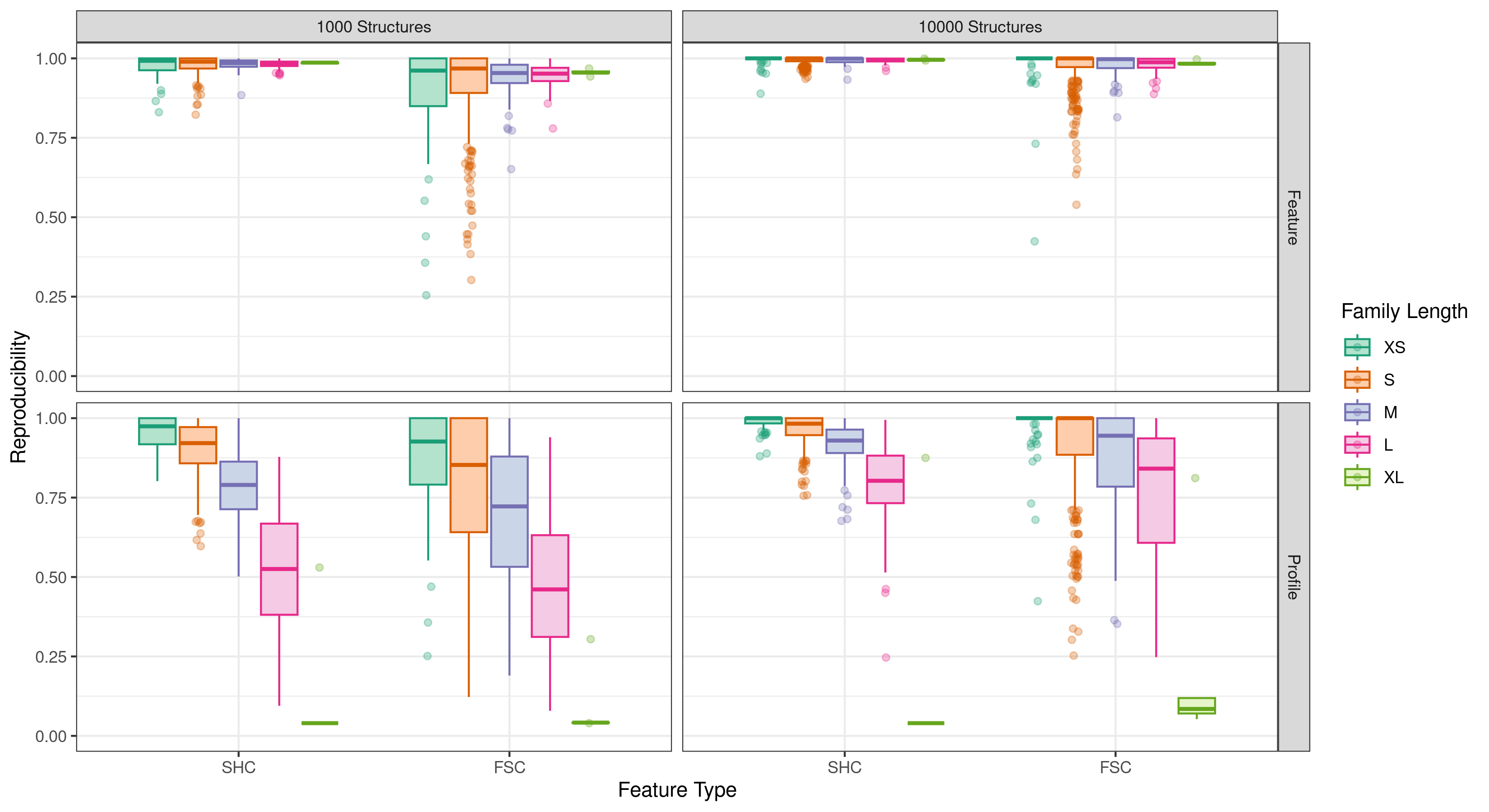}
    \caption{Reproducibility of features and of profiles for 1K and 10K samples across family length categories.
Although 1K suffices for features, profile reproducibility degrades with sequence length.
Larger, e.g.\ 10K, samples yield a more reliable structural signal in general.
	}
    \label{fig:supp_reproducibility}
\end{figure}

Figures~\ref{fig:supp_feature_count},~\ref{fig:supp_profile_length},
and~\ref{fig:supp_reproducibility} provide further details on 
the growth rate of features, growth rate of profile length,
and reproducibility for 1K and 10K samples respectively.

Recall that \vtwo\ denotes RNAprofiling~2.0, and \vone\ the original, 
i.e.\ RNAprofiling~1.0.
Features considered are helices, helices classes (\hc), 
selected helix classes (\shc), stem classes (\stem), and fuzzy 
stem classes (\fsc).
As features,
the difference between \stem\ and \fsc\ is their frequency, 
i.e.\ estimated probability, in the Boltzmann sample input.
As discussed below in Section~\ref{sec:techdet}, this changes  
both the profile selection and the decision tree construction.

\section{Implementation information}

\begin{table}[t]
    \centering
    \begin{tabular}{l|l|l}
        Parameter & Value & Default \\
        \hline
	Output\_Type & \{Hasse, Tree\} & Tree \\
	Feature\_Type & \{Selected Helix Class, Stem Class\} & Stem Class \\
        Frequency\_Format & \{Counts, Percentages, Decimals\} & Counts \\
	Helix\_Class\_Selection\_Cutoff & Positive Integer or Auto & Auto \\ 
	Profile\_Selection\_Cutoff & Positive Integer or Auto & Auto \\ 
\hline
        Stem\_Gap & Non-Negative Integer & 2 \\
	Use\_Fuzzy\_Stem\_Counts & Boolean & True \\
        Fuzzy\_Dilation\_Size & Positive Integer & 5 \\ 
	Fuzzy\_Basepair\_Frequency\_Margin & Real in [0,1] & 0.333 \\ 
	Min\_Contingency\_Node\_Proportion & Real in [0,1] & 0.75 \\
\hline
        Use\_Consistent\_Helix\_Class\_Labels & Boolean & False \\
    \end{tabular}
    \caption{Main code parameters. 
``Cutoff'' options override the standard profiling threshold method;
if used, helix classes, resp.\ profiles, with lower frequency in the
input sample will not be considered.
The three ``Fuzzy'' options apply only to \stem\ and are ignored if using \hc.
``Contingency'' option is ignored if output is not a decision tree.
Consistency in helix labeling can be very useful when comparing multiple
analyses for the same sequence.
}
    \label{tab:params}
\end{table}

\vtwo\ is freely available under the GPLv2 license at 
\texttt{github.com/gtDMMB/RNAprofilingV2}
and can be run online via the 
\texttt{rnaprofiling.gatech.edu} website.

When run online, the default option is to upload a Boltzmann sample
in either ct or dot file format.
To expedite exploratory analysis,
the website also provides the option of generating a sample
with either 
RNAstructure 6.4~\cite{Reuter2010} or ViennaRNA 2.4.14~\cite{Lorenz2011}.

A command line interface is available in the form of a python script.  
The script provides all the same options as the web interface, although 
generating samples is disabled unless there is a local install of 
RNAstructure or ViennaRNA.

This new version has a completely new codebase, 
written in Python rather than C/C++. 
\vtwo\ is implemented  and tested with Python~3 (3.6.9) using the 
numpy (1.19.5), networkx (2.4), 
matplotlib (3.3.4), and pygraphviz (1.6) libraries. 
A graphviz install~\cite{Gansner1999} is also required to generate the 
summary profile graph. 
The program is loaded on the web server using PyInstaller (4.10). 

The full output from both the web interface and the command line interface 
is displayed in HTML with JavaScript and should work in 
most modern internet browsers. 
The svg.min library is used to render graphviz output in the browser. 

Table~\ref{tab:params} lists the main \vtwo\ parameters.  
There are some additional IO options for sequences and samples available
via the command line and website.

\vone\ functionality may be recreated by using \shc\ as features, 
and a Hasse diagram as the summary profile graph output.
By default, a maximum average entropy threshold determines the 
selection of \hc\ and of profiles.
As in \vone, this can be overridden by a user-specified cutoff.

The default options for stem gap, fuzzy counts, and contingency nodes
can be altered by users to suit their particular analysis goals.
For example, by default, 75\% of the full binary tree must be present
before it is collapsed into a contingency node, but can be changed
to provide a larger or smaller output tree as useful.

Finally, we highlight the possibility of having consistent \hc\ labels.
This can be very useful if 
comparing results across multiple different samples for the same sequence.
When invoked, the \hc\ labels are based on the sequence itself, 
and so are independent of the particular sample frequencies.

\clearpage
\pagebreak

\section{Technical details of method}\label{sec:techdet}

\subsection{Stem class length and width}

A helix $(i,j,k)$ has length $k$, the number of base pairs it contains.
A \hc\ is denoted by its maximal helix $(i,j,k)$ and has length $k$
since the maximum number of base pairs possible in any of its
constituent helices is $k$.
Observe that $k - 1$ is half
the Manhattan distance from the outermost possible pairing $(i,j)$
to the innermost $(i+k-1, j-k+1)$ in the usual $(x,y)$ plane;
$k - 1 = (1/2) * (\abs{i - (i+k-1)} + \abs{j - (j-k+1)})$.

The \stem\ length is defined analogously and will be the
maximum number of pairings possible in any of its constituent combinations.
First observe that the outermost possible base pair is
the one with the greatest contact distance, i.e.\ where $j - i$
is maximal. 
Let $M$ denote this value, and $m$ denote least possible,
which corresponds to the innermost pair.
Since the Manhattan distance between the outermost and innermost pairs
is equal to $M - m$,
the length of a stem class is defined to be $1 + (M - m)/2$.

The \stem\ width will capture a measure of the `spread'
of observed pairings represented.
This is done by counting the number ``helical diagonals'' covered by
the \stem.
Here, a helical diagonal denotes a line in the $(x,y)$ plane with slope $-1$
which intersects the identity $x=y$ at points where $x$ or $x + 1/2$
is a positive integer.
The \hc\ $h = (i,j,k)$ has width 1 since all pairings lie on
a single diagonal with midpoint $x = (i+j)/2$ as the intersection.
If the helix class $h' = (i', j', k')$ is stemmable with $h$,
then $\abs{ (i+j)/2 - (i'+j')/2 } \leq 1$, and the number of
diagonals covered is either $1$, $2$, or $3$ depending on
the asymmetry of the internal loop/bulge separating them,
i.e.  whether
$((j-k+1) - j' -1) - (i' - (i+k-1) -1)$ is $0$, $\pm 1$, or $\pm 2$.
For a stem class \ijkl, let $C$ be the set of helix class midpoints.
Then its width is $1 + 2 * (\max C - \min C)$.

\subsection{Fuzzy stem class frequencies}

Fuzzy counts address low-frequency base pairings, i.e.\ non-\shc\ ones,
that are `close' to a \stem.
These augmented frequencies are distinguished as \fsc.
A secondary structure has both a \stem\ profile 
as well as a (possibly enlarged) \fsc\ one.
The frequency of a profile is the number of structures 
in the input sample with those features (and no more).
The difference between \stem\ and \fsc\ as features affects the
distribution of profile frequencies which is used both for selection,
and also to build the decision tree.

\vtwo\ uses fuzzy counts by default, which are found as follows.
Consider a secondary structure $S$ and its \stem\ profile $P$. 
For each \stem\ not already in $P$, expand its region \ijkl\
slightly and count the non-\shc\ base pairs from $S$ which fall inside.
This number is then compared to a baseline.
If it is high enough, then \ijkl\ is added to the \fsc\ profile for $S$.
A base pair $(x,y)$ falls inside the expanded region 
of \ijkl\ with dilation size $b$ if 
${i - b \leq x < i + k + b}$ and ${j - l - b < y \leq j + b}$.
The baseline is a fixed fraction, by default $(1/3)$, 
of the average number of base pairs in the expanded region 
over all structures with \ijkl\ in their \stem\ profile.

\subsection{Decision tree construction}

A path starting at the root of the tree corresponds to a sequence of choices,
positive and/or negative, on features which results in one or more
selected profiles as a leaf.
As will be explained, the multiplicity is due to the use of
``contingency'' leaf nodes.
By default, fuzzy frequencies are used to build the tree,
but exact ones can be chosen instead.

Every node corresponds to a subset of the Boltzmann sample, 
and is labeled with the number of structures under consideration.
The root node is the full sample input, i.e.\ all nonempty profiles.
Subsequent nodes consist of groups of profiles determined by prior 
decisions on feature inclusion/exclusion.
Edges denote decisions and are labeled with one or more features,
with negative choices indicated by $\neg$.
Nontrivial \stem\ are denoted by letters, and trivial ones by their \shc\
integer index.
Subset sizes are updated after every decision to remove the 
profiles no longer in consideration.

The choice of inclusion/exclusion for each remaining feature
splits the selected profiles currently under consideration.
Features which yield the same split are grouped together into a
common decision for consideration.

A decision is ``forced'' if the other side of the split is empty,
i.e.\ if none of the selected profiles in the current group 
have the opposite choice of feature(s).
There is at most one forced common decision possible, and it has priority.
It often includes multiple features, particularly negative options.
In this case, there is a single down edge from the current node.

Otherwise, there will be two down edges, one for each side of
the split for the chosen decision.
Note that both sides contain at least one selected profile.
In this case, 
the different possible common decisions (which may consist of
a single feature) are considered.
The one which maximizes the Hellinger distance is chosen.

The Hellinger distance~\cite{cam-yang-00} is
computed over discrete probability distributions $p$ and $q$
defined on sample space $\mathcal{S}$ as
\begin{equation}
    D_H(p,q,\mathcal{S})=\sqrt{\frac{1}{2} \sum_{s\in \mathcal{S}}\left(\sqrt{p(s)}-\sqrt{q(s)}\right)^2}.
\end{equation}
It is a measure of the similarity of $p$ and $q$, and achieves a maximum
of $1$, i.e.\ the greatest dissimilarity,
when $p$ and $q$ have disjoint support.

Let $F$ be the set of features corresponding to a common split in
the current selected profiles.
The sample space $\mathcal{S}$ for the Hellinger distance computation
corresponding to $F$
is the set of all possible combinations of the remaining features
after removing prior decisions (corresponding to the path to the current node)
and the feature(s) in $F$.
The discrete distributions $p$ and $q$ are the normalized frequencies
from the two sides of the split;
their support over $\mathcal{S}$ is typically sparse.
The decision chosen is the $F$ where $p$ and $q$ are most dissimilar,
with ties broken by lexicographic ordering.

This process of grouping remaining features into decisions,
and choosing one of them, proceeds down each branch of the tree
until a selected profile is reached.
At this point, the tree is evaluated for contingency nodes.

All the descendants of a non-leaf node are replaced by a contingency 
if two conditions are met (and these are not met by any of its ancestors).
First, at least 75\% of the full binary tree is present.
Second, all paths from the node being evaluated to a leaf descendant
have the same set of common decisions.
If so, then that set is presented in a single contingency table.

In this case, all frequencies in the table are reported,
not just the ones for selected profiles.
The decision edges are collapsed down to a single (dashed) contingency
edge, labeled with all the decisions so condensed.
The resulting contingency node (dashed rectangle) compactly represents 
multiple selected profiles.
Its frequency is updated to include the low frequency structures 
reported in the contingency table.